\documentclass
[10pt,twocolumn,oneside,showpacs,pra,superscriptaddress]{revtex4}%
\usepackage{graphicx}
\usepackage{amsmath}%
\usepackage{amsfonts}%
\usepackage{amssymb}

\begin{document}

\title{Experimental Implementation of Quantum Random Walk Algorithm}
\affiliation{Department of Modern Physics, University of Science and Technology of China,
Hefei, 230027, P.R. China}
\affiliation{Department of Physics, National University of Singapore, Lower Fent Ridge,
Singapore 119260}
\affiliation{Harrison M. Randall Laboratory of Physics, The University of Michigan, Ann
Arbor, Michigan 48109-1120}
\affiliation{Laboratory of Structure of Biology, University of Science and Technology of
China, Hefei, 230027, P.R. China}
\author{Jiangfeng Du}
\email{djf@ustc.edu.cn}
\affiliation{Department of Modern Physics, University of Science and Technology of China,
Hefei, 230027, P.R. China}
\affiliation{Department of Physics, National University of Singapore, Lower Fent Ridge,
Singapore 119260}
\author{Hui Li}
\affiliation{Department of Modern Physics, University of Science and Technology of China,
Hefei, 230027, P.R. China}
\author{Xiaodong Xu}
\affiliation{Harrison M. Randall Laboratory of Physics, The University of Michigan, Ann
Arbor, Michigan 48109-1120}
\author{Mingjun Shi}
\affiliation{Department of Modern Physics, University of Science and Technology of China,
Hefei, 230027, P.R. China}
\author{Jihui Wu}
\affiliation{Laboratory of Structure of Biology, University of Science and Technology of
China, Hefei, 230027, P.R. China}
\author{Xianyi Zhou}
\author{Rongdian Han}
\affiliation{Department of Modern Physics, University of Science and Technology of China,
Hefei, 230027, P.R. China}

\begin{abstract}
The quantum random walk is a possible approach to construct new quantum
algorithms. Several groups have investigated the quantum random walk and
experimental schemes were proposed. In this paper we present the experimental
implementation of the quantum random walk algorithm on a nuclear magnetic
resonance quantum computer. We observe that the quantum walk is in sharp
contrast to its classical counterpart. In particular, the properties of the
quantum walk strongly depends on the quantum entanglement.

\end{abstract}
\maketitle

\section{Introduction}

Since the discovery of the first two quantum algorithms, Shor's factoring
algorithm \cite{1} and Grover's database search algorithm \cite{2}, research
in the new born field of quantum computation exploded \cite{3}. Mainly
motivated by the idea that a computational device based on quantum mechanics
might be (possibly exponentially) more powerful than a classical one \cite{4},
great effort has been done to investigate new quantum algorithms and, more
importantly, to experimentally construct a universal quantum computer.
However, finding quantum algorithms is a difficult task. It has also been
proved extremely difficult to experimentally construct a quantum computer, as
well as to carry out quantum computations.

Recently, several groups have investigated quantum analogues of random walk
algorithms \cite{5,6,7,8,9,10}, as a possible direction of research to adapt
known classical algorithms to the quantum case. Random walks on graphs play an
essential role in various fields of natural science, ranging from astronomy,
solid-state physics, polymer chemistry, and biology to mathematics and
computer science \cite{11}. Current investigations show that quantum random
walks have remarkably different features to the classical counterparts
\cite{5,6,7,8,9,10}. The hope is that a quantum version of the random walk
might lead to applications unavailable classically, and to construct new
quantum algorithms. Indeed, the first quantum algorithms based on quantum
walks with remarkable speedup have been reported \cite{al}. Further,
experimental schemes have also been proposed to implement such quantum random
walks by using an ion trap quantum computer \cite{9} and by using neutral
atoms trapped in optical lattices \cite{10}. Up to now, only three methods
have been used to demonstrate quantum logical gates: trapped ions \cite{12},
cavity QED \cite{13} and NMR \cite{14}. Of these methods, NMR has been the
most successful with realizations of quantum teleportation \cite{15}, quantum
error correction \cite{16}, quantum simulation \cite{17}, quantum algorithms
\cite{18}, quantum games \cite{19} and others \cite{20}. In this paper, we
present the experimental implementation of the quantum random walk algorithm
on a two-qubit NMR quantum computer, and we believe it is the first
experimental implementation of such quantum algorithms.

We consider continuous-time random walks (CTRW) \cite{6} rather than
discrete-time random walks \cite{7}, on a circle with four nodes. We show that
the evolution of this quantum walk is periodic and reversible, and yields an
exactly uniform probability distribution at certain time. While the classical
CTRW is irreversible and only approximates the uniform distribution at
infinite-time limit. Further, we experimentally implement this quantum walk on
a two-qubit quantum computer, using a unitary operator which has the right
``effective'' Hamiltonian, with good agreements between theory and experiment.
It is interestingly found that the property of the quantum walk strongly
depends on the entanglement between the two qubits. The uniform distribution
could be obtained only when the two qubits are maximally entangled.

\section{Quantum CTRW on a Circle}

The concept of quantum CTRW is proposed in Ref. \cite{6}. On a circle with
four nodes, we denote the set of nodes by $\left\{  0,1,2,3\right\}  $. Since
the structure of the circle is periodic, only two nodes, $\left(  k+1\right)
\operatorname{mod}4$ and $\left(  k-1\right)  \operatorname{mod}4$, are
connected to node $k$ $\left(  k=0,1,2,3\right)  $. In the classical CTRW, let
$\gamma$ denote the jumping rate, which is a fixed, time-independent constant
that represents the probability moving from a given node to a connected one
per unit time. The generator matrix \cite{6} of this walk can therefore be
written as%
\begin{equation}
H=\left(
\begin{array}
[c]{cccc}%
2\gamma & -\gamma & 0 & -\gamma\\
-\gamma & 2\gamma & -\gamma & 0\\
0 & -\gamma & 2\gamma & -\gamma\\
-\gamma & 0 & -\gamma & 2\gamma
\end{array}
\right)  . \label{eq 1}%
\end{equation}
Consider a particle walking classically on this circle, let $P_{k}^{C}\left(
t\right)  $ denote the probability of being at node $k$ at time $t$ (the
superscript $C$ denotes ``\textit{Classical}''), then we have%
\begin{equation}
\frac{d}{dt}P_{k}^{C}\left(  t\right)  =-\sum\nolimits_{l=0}^{3}H_{kl}%
P_{l}^{C}\left(  t\right)  . \label{eq 2}%
\end{equation}
This equation conserves the probability in the sense that $\sum_{k=0}^{3}%
P_{k}^{C}\left(  t\right)  \equiv1$.

A natural way to the quantum version of this CTRW is to construct a Hilbert
space spanned by the four basis $\left\{  \left\vert 0\right\rangle
,\left\vert 1\right\rangle ,\left\vert 2\right\rangle ,\left\vert
3\right\rangle \right\}  $, respectively corresponding to the four nodes
$\left\{  0,1,2,3\right\}  $. The state of a particle walking quantum
mechanically on this circle is then denoted by $\left\vert \psi\left(
t\right)  \right\rangle $, which is generally a superposition of the four
basis, rather than classical mixing of the probabilities of being at the four
nodes. The generator matrix in classical walks is now treated as the
Hamiltonian of the quantum evolution \cite{6}. Therefore the Schr\"{o}dinger
equation of the state $\left\vert \psi\left(  t\right)  \right\rangle $ is%
\begin{equation}
\frac{d}{dt}\left\langle k\right\vert \left.  \psi\left(  t\right)
\right\rangle =-i\sum\nolimits_{l=0}^{3}\left\langle k\right\vert H\left\vert
l\right\rangle \left\langle l\right\vert \left.  \psi\left(  t\right)
\right\rangle . \label{eq 3}%
\end{equation}
If measuring at time $t$, we can obtain a certain probability distribution on
the circle. The probability of being at node $k$ is $P_{k}^{Q}\left(
t\right)  =\left\vert \left\langle k\right\vert \left.  \psi\left(  t\right)
\right\rangle \right\vert ^{2}$ (the superscript $Q$ denotes \textquotedblleft
\textit{Quantum}\textquotedblright), and the conservation of probability is
obviously guaranteed by the normalization $\left\langle \psi\left(  t\right)
\right\vert \left.  \psi\left(  t\right)  \right\rangle \equiv1$.%

\begin{figure}
[ptb]
\begin{center}
\includegraphics[
height=5.3334cm,
width=7.3323cm
]%
{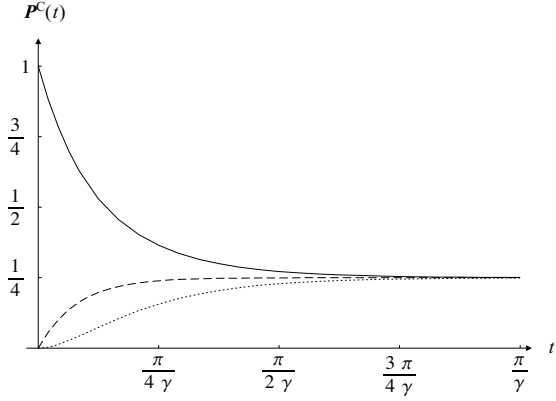}%
\caption{The probabilities of being at the four nodes in the classical version
of this CTRW. The solid (dotted) line corresponds to $P_{0}^{C}\left(
t\right)  $ ($P_{2}^{C}\left(  t\right)  $). The dashed line corresponds to
$P_{1}^{C}\left(  t\right)  $ and $P_{3}^{C}\left(  t\right)  $, since
$P_{1}^{C}\left(  t\right)  =P_{3}^{C}\left(  t\right)  $.}%
\label{ProbabilityPlotC}%
\end{center}
\end{figure}

Let the walking particle start from node $0$, it is then easy to find the
probability of being at node $k$ at any time $t$ in both the classical walks
and the quantum walks (by solving equations (\ref{eq 2}) and (\ref{eq 3}),
respectively). The detailed calculation yields that for the classical walks
the probabilities are%
\begin{equation}
\left\{
\begin{array}
[c]{l}%
P_{0}^{C}\left(  t\right)  =\frac{1}{4}+\frac{1}{2}e^{-2\gamma t}+\frac{1}%
{4}e^{-4\gamma t}\\
P_{1}^{C}\left(  t\right)  =P_{3}^{C}\left(  t\right)  =\frac{1}{4}-\frac
{1}{4}e^{-4\gamma t}\\
P_{2}^{C}\left(  t\right)  =\frac{1}{4}-\frac{1}{2}e^{-2\gamma t}+\frac{1}%
{4}e^{-4\gamma t}%
\end{array}
\right.  . \label{eq 4}%
\end{equation}%

\begin{figure}
[ptb]
\begin{center}
\includegraphics[
height=2.0998in,
width=2.8867in
]%
{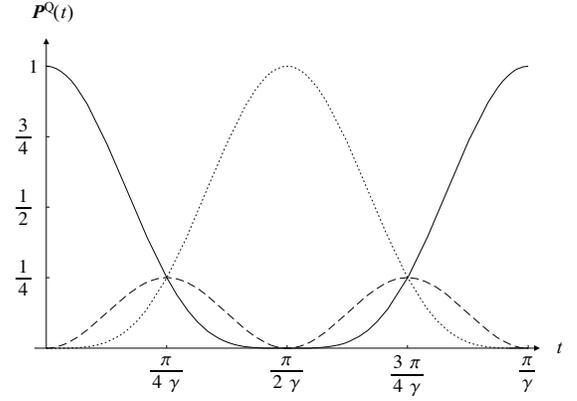}%
\caption{The probabilities of being at the four nodes in the quantum version
of this CTRW. The solid (dotted) line corresponds to $P_{0}^{Q}\left(
t\right)  $ ($P_{2}^{Q}\left(  t\right)  $). The dashed line corresponds to
$P_{1}^{Q}\left(  t\right)  $ and $P_{3}^{Q}\left(  t\right)  $, since
$P_{1}^{Q}\left(  t\right)  =P_{3}^{Q}\left(  t\right)  $.}%
\label{ProbabilityPlotQ}%
\end{center}
\end{figure}

In the quantum case, the initial state of the particle is $\left\vert
\psi\left(  0\right)  \right\rangle =\left\vert 0\right\rangle $, from
equation (\ref{eq 3}), we have
\begin{align}
\left\vert \psi\left(  t\right)  \right\rangle  &  =e^{-iHt}\left\vert
0\right\rangle \nonumber\\
&  =e^{-2i\gamma t}\cos^{2}\gamma t\left\vert 0\right\rangle -e^{-2i\gamma
t}\sin{}^{2}\gamma t\left\vert 2\right\rangle \nonumber\\
&  +\frac{i}{2}e^{-2i\gamma t}\sin2\gamma t\left(  \left\vert 1\right\rangle
+\left\vert 3\right\rangle \right)  . \label{eq 6}%
\end{align}
Therefore the probabilities of the quantum walks are%
\begin{equation}
\left\{
\begin{array}
[c]{l}%
P_{0}^{Q}\left(  t\right)  =\cos^{4}\gamma t\\
P_{1}^{Q}\left(  t\right)  =P_{3}^{Q}\left(  t\right)  =\frac{1}{4}\sin
^{2}2\gamma t\\
P_{2}^{Q}\left(  t\right)  =\sin{}^{4}\gamma t
\end{array}
\right.  . \label{eq 5}%
\end{equation}

The probabilities in equations (\ref{eq 4}, \ref{eq 5}) are plotted in FIG.
\ref{ProbabilityPlotC} (for the classical walks) and FIG.
\ref{ProbabilityPlotQ} (for the quantum walks) as functions of time $t$. From
FIG. \ref{ProbabilityPlotC} and FIG. \ref{ProbabilityPlotQ}, we can see
striking differences between quantum and classical random walks. FIG.
\ref{ProbabilityPlotQ} shows that the evolution of the quantum CTRW on a
circle with four node are essentially periodic with a period $T=\pi/\gamma$.
The particle walking quantum mechanically on this circle will definitely go
back to its original position, and the evolution is reversible. It is also
interesting to see that at time $t=\pi/2\gamma$ the probability distribution
converges to node $2$. These phenomena are due to the quantum interference
effects, which allows probability amplitudes from different paths to cancel
each other.

\begin{figure}
[ptb]
\begin{center}
\includegraphics[
height=1.913in,
width=2.8867in
]%
{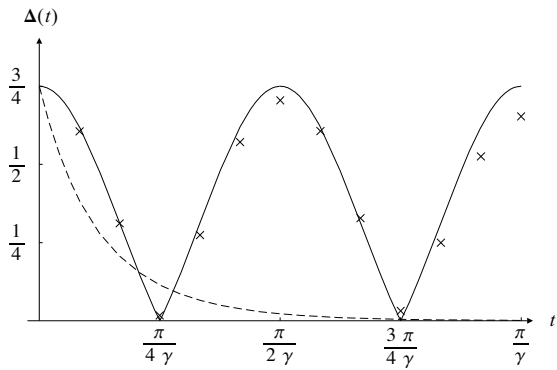}%
\caption{The quantum and classical probability distributions as functions of
time $t$. The solid line corresponds to $\Delta^{Q}\left(  t\right)  $, and
the dashed line to $\Delta^{C}\left(  t\right)  $, both in theory. The crosses
corresponds to the experimental results of the quantum case.}%
\label{DistributePlots}%
\end{center}
\end{figure}

To measure how uniform a distribution is, an immediate way is to use the
\textit{total variation distance} between the given distribution and the
uniform distribution. In our case, the classical and quantum total variation
distance as functions of time $t$ are%
\begin{align}
\Delta^{C}\left(  t\right)   &  =\frac{1}{2}\sum\nolimits_{k=0}^{3}\left\vert
P_{k}^{C}\left(  t\right)  -\tfrac{1}{4}\right\vert ,\label{eq 9}\\
\Delta^{Q}\left(  t\right)   &  =\frac{1}{2}\sum\nolimits_{k=0}^{3}\left\vert
P_{k}^{Q}\left(  t\right)  -\tfrac{1}{4}\right\vert . \label{eq 10}%
\end{align}
FIG. \ref{DistributePlots} depicts the dependence of $\Delta^{C}\left(
t\right)  $ and $\Delta^{Q}\left(  t\right)  $ on time $t$. From FIG.
\ref{DistributePlots}, we can see that the classical version of this walking
process approaches the uniform distribution exponentially as time lapses. In
contrast, the quantum process exhibits an oscillating behavior. An intriguing
property of this quantum random walk is that $\Delta^{Q}\left(  n\pi
/4\gamma\right)  =0$ if $n$ is odd, which means that the probability
distribution is exactly uniform at time $t=\pi/4\gamma$ and its odd multiples.
While the classical walk can never reach the exactly uniform distribution,
only approximates it at infinite-time limit.%

\section{Experimental Implementation}

For the quantum CTRW on a circle with four node, the Hilbert space is
4-demensional. So it is natural to implement the quantum walks on a two-qubit
quantum computer. The direct correspondence is to map the basis \{$\left\vert
0\right\rangle $, $\left\vert 1\right\rangle $, $\left\vert 2\right\rangle $,
$\left\vert 3\right\rangle \}$ of the quantum CTRW into the four computational
basis \{$\left\vert 0\right\rangle \otimes\left\vert 0\right\rangle $,
$\left\vert 0\right\rangle \otimes\left\vert 1\right\rangle $, $\left\vert
1\right\rangle \otimes\left\vert 0\right\rangle $, $\left\vert 1\right\rangle
\otimes\left\vert 1\right\rangle $\}. This mapping is in fact to rephrase the
number of nodes by the binary number system. Therefore the Hamiltonian in
equation (\ref{eq 1}) can be written as%
\begin{equation}
H=2\gamma I\otimes I-\gamma\left(  I\otimes\sigma_{x}+\sigma_{x}\otimes
\sigma_{x}\right)  , \label{eq 7}%
\end{equation}
where $I$ and $\sigma_{x}$ are the identity operator and the Pauli operator of
a single qubit. The evolution operator of the two-qubit system is%
\begin{align}
U\left(  t\right)   &  =e^{-iHt}\nonumber\\
&  =e^{-2i\gamma t}\exp\left[  i\gamma t\left(  \sigma_{x}\otimes\sigma
_{x}\right)  \right]  \exp\left[  i\gamma t\left(  I\otimes\sigma_{x}\right)
\right]  . \label{eq 8}%
\end{align}
And the state of a particle performing this quantum CTRW is%
\begin{align}
\left\vert \psi\left(  t\right)  \right\rangle  &  =e^{-2i\gamma t}\cos
^{2}\gamma t\left\vert 00\right\rangle -e^{-2i\gamma t}\sin{}^{2}\gamma
t\left\vert 10\right\rangle \nonumber\\
&  +\frac{i}{2}e^{-2i\gamma t}\sin2\gamma t\left(  \left\vert 01\right\rangle
+\left\vert 11\right\rangle \right)  . \label{eq 11}%
\end{align}
It is interesting to investigate the relations between the distribution of the
implemented quantum CTRW and the entanglement of the two-qubit state
$\left\vert \psi\left(  t\right)  \right\rangle $. The entanglement of the
two-qubit state $\left\vert \psi\left(  t\right)  \right\rangle $ in equation
(\ref{eq 11}) can be directly calculated by the Von. Neumann entropy%
\begin{equation}
S\left(  t\right)  =-\cos^{2}\gamma t\log_{2}\left(  \cos^{2}\gamma t\right)
-\sin{}^{2}\gamma t\log_{2}\left(  \sin^{2}\gamma t\right)  . \label{eq 12}%
\end{equation}%

\begin{figure}
[ptb]
\begin{center}
\includegraphics[
height=1.8412in,
width=2.885in
]%
{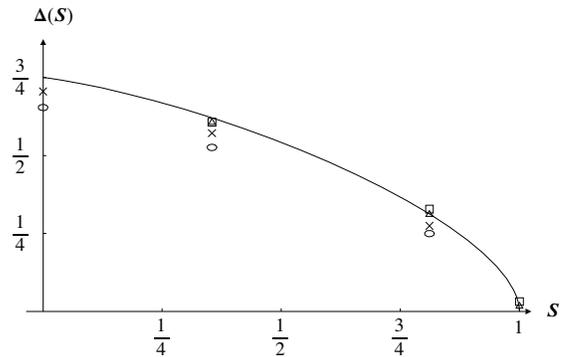}%
\caption{The correlation between the quantum total variation distance
$\Delta^{Q}\left(  t\right)  $ and the entanglement $S\left(  t\right)  $. The
line corresponds to theoretical calculation. The experimental results in
different sets are shown in different shapes. The triangles are for
$n\in\left\{  1,2,3\right\}  $, crosses for $n\in\left\{  4,5,6\right\}  $,
boxes for $n\in\left\{  7,8,9\right\}  $ and circles for $n\in\left\{
10,11,12\right\}  $.}%
\label{EntanglementPlot}%
\end{center}
\end{figure}

The correlation between the quantum total variation distance $\Delta
^{Q}\left(  t\right)  $ and the entanglement $S\left(  t\right)  $ is
illustrated in FIG. \ref{EntanglementPlot}. From FIG. \ref{EntanglementPlot},
we can see that if there is no entanglement between the two qubits ($S=0$),
$\Delta^{Q}$ is at its maximal $\Delta^{Q}=\frac{3}{4}$, which corresponds
converging at node $0$ (or node $2$). While if the two qubits are maximally
entangled ($S=1$), $\Delta^{Q}=0$, which happens to be the situation that the
walking particle is uniformly distributed on the four nodes. Therefore, we can
say that the quantum random walk algorithm is enhanced by the quantum
entanglement involved.

The quantum CTRW is implemented using our two-qubit NMR quantum computer. This
computer uses a $0.5$ milliliter, $200$ millimolar sample of Carbon-13 labeled
chloroform (Cambridge Isotopes) in d$_{\text{6}}$ acetone. In a magnetic
field, the two spin states of $^{\text{1}}$H\ and $^{\text{13}}$C\ nuclei in
the molecular can be described as four nodes of two qubits, while radio
frequency (RF) fields and spin--spin couple constant $J$ are used to implement
quantum network of CTRW. Experimentally, we performe twelve separate sets of
experiments with various selection of time $t$ which is distinguished by
$\gamma t=\frac{n\cdot\pi}{12}\left(  n=\left\{  0,1,2,\cdots,12\right\}
\right)  $. In the following, we replace the jumping rate $\gamma$ with $\pi
J$ $(J=215Hz)$\textit{.} In each set, the full process of the quantum CTRW is
executed. We describe this experimental process as follows.

Firstly, prepare effective pure state $\left|  \psi\left(  0\right)
\right\rangle $: The initial state in NMR is thermally equilibrium state
$\rho_{eq}\equiv4I_{z}^{1}+I_{z}^{2}$ rather than a true pure state $\left|
\psi\left(  0\right)  \right\rangle $. However, it is possible to create an
effective pure state, which behaves in an equivalent manner. This is
implemented as
\[
R_{x}^{1}\left(  \pi/3\right)  -G_{z}-R_{x}^{1}\left(  \pi/4\right)
-\tau-R_{y}^{1}\left(  -\pi/4\right)  -G_{z},
\]
to be read from left to right, radio-frequency pulses are indicated by
$R_{axis}^{spins}\left(  angle\right)  $, and are applied to the spins in the
superscript, along the axis in the subscript, by the angle in the brackets.
For example, $R_{x}^{1}\left(  \pi/3\right)  $ denotes $\pi/3$ selective pulse
that acts on the first qubit about $\widehat{x}$, and so forth. $G_{z}$ is the
pulsed field gradient along the $\widehat{z}$ axis to annihilate transverse
magnetizations, dashes are for readability only, and $\tau$ represents a time
interval of $1/\left(  2J\right)  $. Therefore, after the state preparation,
we obatin effective pure state $\rho\left(  0\right)  \equiv I_{z}^{1}%
+I_{z}^{2}+2I_{z}^{1}I_{z}^{2}$ from equilibrium state $\rho_{eq}\equiv
4I_{z}^{1}+I_{z}^{2}$.

Secondly, perform quantum CTRW with different time $t$: As shown above in
equation (\ref{eq 8}), quantum CTRW can be described as unitary operator
$U\left(  t\right)  $, this is performed with pulse sequence shown in the
following (Note that the global phase $e^{-2i\gamma t}$ of $U\left(  t\right)
$ is safely ignored in our experiments, since $\rho\left(  t\right)  =U\left(
t\right)  \rho\left(  0\right)  U\left(  t\right)  ^{\dag}$, this global phase
is meaningless and has no effect on the result of experiment)%
\[
R_{x}^{2}\left(  \theta\right)  -R_{y}^{12}\left(  \pi/2\right)  -\frac{t}%
{2}-R_{x}^{12}\left(  \pi\right)  -\frac{t}{2}-R_{y}^{12}\left(
-\pi/2\right)  .
\]
Here $R_{x}^{2}\left(  \theta\right)  $ is equal to $e^{-i\frac{\theta}%
{2}\widehat{\sigma}_{x}}$ that act on the second spin,where $\theta=n\pi/6$
and $t=n/\left(  6J\right)  =n\pi/\left(  6\gamma\right)  $ for $n\in\left\{
1,2,\cdots,12\right\}  $, $R_{x}^{12}\left(  \pi\right)  $ denotes $\pi$
non-selective pulse that acts on both qubits about $\widehat{x}$. It is
obviously that the final state $\rho\left(  t\right)  $ of the quantum CTRW
prior to measure is given by $\rho\left(  t\right)  =U\left(  t\right)
\rho\left(  0\right)  U\left(  t\right)  ^{\dag}$.

Finally, readout the result $\rho\left(  t\right)  $ and calculate quantum
total variation distance $\Delta^{Q}\left(  t\right)  $: In NMR experiment, it
is not practical to determine the final state directly, but an equivalent
measurement can be made by so-called quantum state tomography to recover the
density matrix $\rho\left(  t\right)  =\left|  \psi\left(  t\right)
\right\rangle \left\langle \psi\left(  t\right)  \right|  $. However, as only
the diagonal elements of the final density operators are needed in our
experiments, the readout procedure is simplified by applying gradient pulse
before readout pulse to cancel the off-diagonal elements. Here we shall note
that, the gradient pulse can remove off-diagonal terms since we use
heteronuclear systems in our experiment. Then quantum total variation distance
$\Delta^{Q}\left(  t\right)  $ is determined by the equation $\Delta
^{Q}\left(  t\right)  =\tfrac{1}{2}%
{\textstyle\sum\nolimits_{k=0}^{3}}
\left|  P_{k}^{Q}\left(  t\right)  -\tfrac{1}{4}\right|  $, where $P_{k}%
^{Q}\left(  t\right)  =\left\langle k\right|  \rho\left(  t\right)  \left|
k\right\rangle \ $\ is certain probability on the node $\left|  k\right\rangle
$. Finally, $\Delta^{Q}\left(  S\right)  $ is determined with equation
(\ref{eq 12}).

All experiments are conducted at room temperature and pressure on Bruker
Avance DMX-500 spectrometer in Laboratory of Structure Biology, University of
Science and Technology of China. FIG. \ref{DistributePlots} show the quantum
total variation distance $\Delta^{Q}\left(  t\right)  $ as a function of time
$t$ and FIG. \ref{EntanglementPlot} show the quantum total variation distance
$\Delta^{Q}\left(  S\right)  $ as a function of entanglement $S$ of
$\left\vert \psi\left(  t\right)  \right\rangle $ shown in equation (\ref{eq
11}). From FIG. \ref{DistributePlots} and FIG. \ref{EntanglementPlot}, it is
clearly seen the good agreement between theory and experiment. However, there
exist small errors which increase when time $t$ increase, we think that the
most errors are primarily due to decoherence, because the time used to
implement quantum CTRW $U\left(  t\right)  $ is increased from several to
several tens milliseconds approximately, while the decoherence time
$T_{2}\approx0.3$ and $0.4$ seconds for carbon and proton respectively. The
other errors are due to inhomogeneity of magnetic field, imperfect pulses, and
the variability over time of the measurement process.

\section{Conclusion}

We present the experimental implementation of the quantum random walk
algorithm on a two-qubit nuclear magnetic resonance quantum computer. For the
quantum CTRW on a circle with four nodes, we observe that the quantum walk
behaves greatly differently from its classical version. The quantum CTRW can
yield an exactly uniform distribution, and is reversible and periodic, while
the classical walk is essentially dissipative. Further, we find that the
property of this quantum walk strongly depends on the quantum entanglement
between the two qubits. The uniform distribution could be obtained only when
the two qubits are maximally entangled. In this paper only the relatively
simple case with two qubits are considered. However, our scheme could be
extended to the case of a graph containing arbitrary $N$ nodes, and the
quantum random walk could be carried out by using $\log_{2}N$ qubits.

\begin{acknowledgments}
We thank T. Rudolph and Y.D. Zhang for useful discussion. Part of the ideas
were originated while J.F. Du was visiting Service de Physique Th\'{e}orique,
Universit\'{e} Libre de Bruxelles, Bruxelles. S. Massar and N. Cerf are
gratefully acknowledged for their invitation and hospitality. J.F. Du also
thanks Dr. J.F. Zhang for the loan of the chloroform sample. This work was
supported by the National Nature Science Foundation of China (Grants No.
10075041 and No. 10075044), the National Fundamental Research Program
(2001CB309300) and the ASTAR Grant No. 012-104-0040.
\end{acknowledgments}


\begin{thebibliography}{9}                                                                                                %

\bibitem {1}P.W. Shor, in \textit{Proceedings of the 35th Annual Symposium on
Foundations of Computer Science} (IEEE Computer Society Press, Los Alamitos,
CA, 1994), p. 124.

\bibitem {2}L.K. Grover, Phys. Rev. Lett. \textbf{79}, 325 (1997).

\bibitem {3}N.A. Nielsen and I.L. \textit{Chuang, Quantum Computation and
Quantum Information} (Cambridge University Press, Cambridge, 2000).

\bibitem {4}R.P. Feynman, Int. J. Theor. Phys. \textbf{21}, 467 (1982).

\bibitem {5}Y. Aharonov, \textit{et al.}, Phys. Rev. A \textbf{48}, 1687 (1993).

\bibitem {6}E. Farhi and S. Gutmann, Phys. Rev. A \textbf{58}, 915 (1998).

\bibitem {7}A. Ambainis, \textit{et al.}, in \textit{Proceedings of the 30th
annual ACM Symposium on Theory of Computing} (Association for Computing
Machinery, New York, 2001).

\bibitem {8}A.M. Childs \textit{et al.}, quant-ph/0103020; C. Moore and A.
Russell, quant-ph/0104037; N. Konno \textit{et al.}, quant-ph/0205065; J.
Kempe, quant-ph/0205083; T. Yamasaki \textit{et al.}, quant-ph/0205045; E.
Bach \textit{et al.}, quant-ph/0207008.

\bibitem {9}B.C. Travaglione and G.J. Milburn, Phys. Rev. A \textbf{65},
032310 (2002).

\bibitem {10}W. D\"{u}r \textit{et al.}, Phys. Rev. A \textbf{66}, 052319 (2002).

\bibitem {11}M.N. Barber and B.W. Ninham, \textit{Random and Restricted Walks:
Theory and Applications} (Gordon and Breach, New York, 1970).

\bibitem {al}A.M. Childs \textit{et al.}, quant-ph/0209131; N. Shenvi
\textit{et al.}, quant-ph/0210064.

\bibitem {12}J.I. Cirac and P. Zoller, Phys. Rev. Lett. \textbf{74}, 4091 (1995).

\bibitem {13}C. Monroe \textit{et al.}, Phys. Rev. Lett. \textbf{75}, 4714 (1995).

\bibitem {14}D.G. Cory, A.F. Fahmy, and T.F. Havel, Proc. Natl. Acad. Sci. USA
\textbf{94}, 1634 (1997); N. Gershenfeld and I. Chuang, Science \textbf{275},
350 (1997).

\bibitem {15}M.A. Nielsen, E. Knill and R. Laflamme, Nature \textbf{396}, 52 (1998).

\bibitem {16}D.G. Cory \textit{et al.}, Phys. Rev. Lett. \textbf{81}, 2152
(1998); D. Leung \textit{et al.}, Phys. Rev. A \textbf{60}, 1924 (1999); E.
Knill \textit{et al.}, Phys. Rev. Lett. \textbf{86}, 5811 (2001).

\bibitem {17}S. Somaroo \textit{et al.}, Phys. Rev. Lett. \textbf{82}, 5381 (1998).

\bibitem {18}L.M.K. Vandersypen \textit{et al., }Nature\ \textbf{414}, 883
(2002); I.L. Chuang \textit{et al.}, Nature\ \textbf{393}, 143 (1998); J.A.
Jones, M. Mosca and R.H. Hansen, Nature \textbf{393}, 344 (1998); R. Marx
\textit{et al.}, Phys. Rev. A \textbf{62},\ 012310 (2000); J. Du \textit{et
al.}, Phys. Rev. A \textbf{64},\ 042306 (2001). L.Xiao \textit{et al.}, Phys.
Rev. A \textbf{66},\ 052320 (2002). X. Peng \textit{et al.}, Phys. Rev. A
\textbf{65},\ 042315 (2002). J. Zhang \textit{et al.}, Phys. Rev. A
\textbf{66},\ 044308 (2002).

\bibitem {19}J. Du \textit{et al., }Phys. Rev. Lett. \textbf{88}, 137902 (2002).

\bibitem {20}X. Fang \textit{et al.}, Phys. Rev. A \textbf{61},\ 022307
(2000); J. Du \textit{et al.}, Phys. Rev. A \textbf{63},\ 042302 (2001); D.
Collins \textit{et al.}, Phys. Rev. A \textbf{62},\ 022304 (2000).
\end{thebibliography}
\end{document}